# IP Tracing and Active Network Response

Tarek S. Sobh
Egyptian Armed Forces, Cairo, Egypt
tarekbox2000@arabia.com

Awad H. Khalil
Department of Computer Science,
The American University in Cairo, Egypt
akhalil@aucegypt.edu

## Abstract

Active security is mainly concerned with performing one or more security functions when a host in a communication network is subject to an attack. Such security functions include appropriate actions against attackers. To properly afford active security actions a set of software subsystems should be integrated together so that they can automatically detect and appropriately address any vulnerability in the underlying network.

This work presents integrated model for active security response model. The proposed model introduces Active Response Mechanism (ARM) for tracing anonymous attacks in the network back to their source. This work is motivated by the increased frequency and sophistication of denial-of-service attacks and by the difficulty in tracing packets with incorrect, or "spoofed", source addresses. This paper presents within the proposed model two tracing approaches based on:
- Sleepy Watermark Tracing (SWT) for unauthorized access attacks.
- Probabilistic Packet Marking (PPM) in the network for Denial of Service (DoS) and Distributed Denial of Service (DDoS) attacks.

On the basis of the proposed model a cooperative network security tools such as firewall, intrusion detection system with IP tracing mechanism has been designed for taking a rapid active response against real IPs for attackers. The proposed model is able to detect network vulnerabilities, trace attack source IP and reconfigure the attacked subnetworks.

**Keywords** - IP Tracing, Sleepy Watermark Tracing, Probabilistic Packet Marking, Active Security, Intrusion Detection System,

## 1. Introduction

Network attacks such as Denial-of-Service (DoS) attack consume the resources of a remote host or network, thereby denying or degrading service to legitimate users. The main aim of a DoS is the disruption of services by attempting to limit access to a machine or service instead of subverting the service itself. This kind of attack aims at rendering a network incapable of providing normal service by targeting either the network's bandwidth or its connectivity. Such attacks are among the hardest security problems to address because they are simple to implement, difficult to prevent, and very difficult to trace.

Intrusion Detection Systems (IDS) attempts to detect intrusion through analysing observed system or network activities [Oh 2003]. Based on the type of observed activities, IDS can be classified as network-based or host-based. IDS will raise alarms when it has detected misuse or anomaly. It may also report intrusion by emailing or paging system administrator and even disconnected intrusion connection locally. However, IDS is passive and it gives little information about where the intrusion really comes from.

Unfortunately, mechanisms for dealing with network attacks have not advanced at the same pace. Most work in this area has focused on tolerating attacks by reducing their effects on the victim [Spatscheck 1999][Banga 1999][Karn 1999]. This approach can provide an effective stop-gap measure, but does not eliminate the problem nor does it discourage attackers. The other option, and the focus of this paper, is to present active network model to trace attacks back to their origin then take appropriate action against attacker so they can be eliminated at the source.

One of the biggest problems in computer security is "DDoS" Distributed Denial of Service. To take an important step towards the solution of such problem, the active network protection approach, in which not only the intruder is detected but also an action must be taken against it. Determining the source of an attack, which is called the traceback problem [Song 2000], is important for such active network protection approach. But it is surprisingly difficult due to the stateless nature of Internet routing. Attackers routinely disguise their location using incorrect, or "spoofed", IP source addresses. As these packets traverse the Internet their true origin is lost and a victim is left with little useful information [Savage 2000]. While there are several tracing techniques in use, they all do not integrate with a complete active network model.

This paper is organized as follows: In Section 2, we introduce a classification of existing tracing approaches. In Section 3 we describe existing schemes for combating anonymous attacks. Section 4 contains a description for the proposed model. Finally, we summarize our findings in Section 5.

## 2. Tracing Approaches Classifications

In general, tracing approaches for a connection chain can be divided into two categories: host-based and network-based, each of which can further be classified into either active or passive. Table (1) provides a classification of existing tracing approaches.

**Table (1):** Classification of Existing Tracing Approaches and Sleepy Watermark Tracing (SWT) [Wang 2001]

|  | **Passive** | **Active** |
|---|---|---|
| **Host-based** | DIDS<br>CIS | Caller ID |
| **Network-based** | Thumbprinting<br>Timing-Based<br>Deviation-based | IDIP<br>SWT |

Host-based tracing approaches depend on information collected from each host involved in the tracing. Distributed Intrusion Detection System (DIDS) attempts to keep track of all the users in the network and account for all activities to network-wide intrusion detection systems [Oh 2003]. Each monitored host in the DIDS domain collects audit trails and sends audit abstracts to a centralized DIDS director for analysis.

The caller Identification System (CIS) by [Jung 1993] is anther host-based tracing mechanism. Each host along the chain keeps a record about its view of the login chain so far. When the user from the n-/th host attempts to login into the nth host, the nth host asks the n-1th host about its view of the login chain of that user, which should be 1,2 …n-1 ideally. The nth host then queries host n-1 to 1 about their views of the login chain as so on. Only when the login chain information from all queried hosts matches, will the login be granted at nth host.

Caller ID, described by [Staniford-Chen 1995], is yet another interesting host-based approach that is said to be used by the Air Force. Caller ID is controversial in that it actually utilizes the same break-in technique used by intruders to break into hosts along the connection chain reversibly. If the intruder from Hi connects through H₁, H2…Hn-1 to the final target Hn, the network security personnel at Hn first breaks into Hn-1; from there they can find out the intruder comes from Hn-2, then they break into Hn-2 and so on. Eventually they can find the origin of the intruder.

The fundamental problem with the host-based tracing approach is its trust model. Host-based tracing places its trust upon the monitored hosts themselves. In specific, it depends on the correlation of connections at every host in the connection chain. If one host is compromised and is providing misleading correlation information, the whole tracing system is fooled. Because host-based tracing requires participation and trust of every host involved in the network-based intrusion, it is very difficult to be applied in the context of the public Internet.

Network-based tracing is the other category of tracing approaches. Neither does it require the participation of monitored hosts, nor does it place its trust on the monitored hosts. It is based on the property of network connections: the application level content of chained connections is invariant across the connection chain. In particular, the thumbprint [Staniford-Chen 1995] is a pioneering correlation technique that utilizes a small quantity of information to summarize connections. Ideally it can uniquely distinguish a connection from unrelated connections and correlate those related connections in the same connection chain. While thumb printing can be useful even when only part of the Internet implements it, it depends in clock synchronization to match thumbprints of corresponding intervals of connections. It also is vulnerable to retransmission variation. This severely limits its usefulness in real-time tracing.

Active network protection approach must be taken against intruder, must be followed.
- In order to take an appropriate action against the attacker, it must be identified, in other words, whether the attacking IP is a *masked IP* or it is a *real IP*.
- Distinguishing Between the masked IP and the Real IP helps us to take the appropriate action against the proper attacker not the wrong one.
- Many Approaches can be used to identify the real IP from the masked IP such as watermark tracing approach and statistical approach.

**Sleepy Watermark Tracing (SWT)** is a way to address the problem of network-based intrusion. In other words, SWT has the ability to effectively trace the detected intrusion that utilizes stepping stone to disguise its origin at real-time, and dynamically push the intrusion countermeasures such as remote monitoring, blocking, containment and isolation close to the source of the intrusion [Wang 2001].

SWT identified the following assumptions that motivate and constrain the design:
- Intrusions are interactive and bi-directional,
- Routers are trust worthy and hosts are not trust worthy,
- Each host a single SWT guarding gateway and
- There is no link-to-link encryption.

The first two assumptions represent the assessments of the nature of intrusions. *In SWT approach, intrusions are those attacks aiming to gain unauthorized access, rather than denial of service attacks*. The compromised router needs to be addressed first, before the tracing of the intrusion can go any further. The assumption of each host having a single SWT guardian gateway is only for simplifying, in case some host has multiple SWT guardian gateways, the guardian gateway set will be used in SWT tracing. The final assumption represents the inherent limitation of any tracing based on network content. SWT supposes a correlation of encrypted connections in real-time is still an open problem.

Without effective intrusion source tracing, intrusion response is limited to the nearby of intrusion target and is passive in front of network-based intrusions. On the other hand, effective intrusion source tracing enables us to build a more active and dynamic intrusion response by pushing the intrusion countermeasures near the source of network-based intrusions.

Automatic and network-wide intrusion responses can be built on SWT guardian gateway by applying active network principles [Wang 2001]:
- Remote monitoring and surveillance.
- Remote decoy and trap.
- Remote blocking and containment.
- Remote isolation and quarantine.
- Dynamic perimeter defense.

While these examples of active intrusion response are based on SWT. The proposed model uses SWT as a real-time intrusion source tracing.

# 3. Existing Schemes

There have been several efforts to reduce the anonymity afforded by IP spoofing. Table (2) provides a subjective characterization of each of these approaches in terms of management cost, additional network load, overhead on the router, the ability to trace multiple simultaneous attacks, and the ability trace attacks after they have completed.

## 3.1 Ingress filtering

Obviously, the best way to address the problem of anonymous attacks is to eliminate the ability to forge source addresses. One such approach, frequently called ingress filtering, is to configure routers to block packets that arrive with illegitimate source addresses [Ferguson 1998]. This requires a router with sufficient power to examine the source address of every packet and sufficient knowledge to distinguish between legitimate and illegitimate addresses. Consequently, ingress filtering is most feasible at the border of Internet Service Providers (ISP) where address ownership is relatively unambiguous and traffic load is low. As traffic is aggregated from multiple ISPs into transit networks, there is no longer enough information to unambiguously determine if a packet arriving on a particular interface has a "legal" source address. Moreover, on such high speed links the overhead of comparing every packet to a filter list becomes prohibitive.

**Table (2):** Qualitative comparison of existing schemes for combating anonymous attacks [Savage 2000].

|  | Management overhead | Network overhead | Router overhead | Distributed capability | Post-mortem capability |
|---|---|---|---|---|---|
| Ingress filtering | Moderate | Low | Moderate | N/A | N/A |
| Link testing |  |  |  |  |  |
|   Input debugging | High | Low | High | Good | Poor |
|   Controlled flooding | Low | High | Low | Poor | Poor |
| Logging | High | Low | High | Excellent | Excellent |
| Marking | Low | Low | Low | Excellent | Excellent |

The principal problem with ingress filtering is that its effectiveness depends on widespread, if not universal, deployment. Unfortunately, a significant fraction of ISPs, perhaps the majority, do not implement this service – either because they are uninformed or have been discouraged by the administrative burden1, potential router overhead and complications with services like Mobile IP [Perkins 1996]. A secondary problem is that even if ingress filtering were universally deployed at the customer-to-ISP level, attackers could still forge addresses from the hundreds or thousands of hosts within a valid customer network [Ferguson 1998].

## 3.2 Link testing

Most existing traceback techniques start from the router closest to the victim and interactively test its upstream links until they determine which one is used to carry the attacker's traffic. Ideally, this procedure is repeated recursively on the upstream router until the source is reached. This technique assumes that an attack remains active until the completion of a trace and is therefore inappropriate for attacks that are detected after the fact, attacks that occur intermittently, or attacks that modulate their behavior in response to a traceback. There are two varieties of link testing schemes, input debugging and controlled flooding.

### 3.2.1 Input debugging
Some routers include a feature called input debugging that allows an operator to filter particular packets on some egress port and determine which ingress port they arrived on. This capability is used to implement a trace as follows: First, the victim must recognize that it is being attacked and develop an attack signature that describes a common feature contained in all the attack packets. The victim communicates this signature to a network operator, frequently via telephone, who then installs a corresponding input debugging filter on the victim's upstream egress port. This filter reveals the associated input port, and hence which upstream router originated the traffic. The process is then repeated recursively on the upstream router, until the originating site is reached or the trace leaves the ISP's border (and hence its administrative control over the routers).

### 3.2.2 Controlled flooding

Burch and Cheswick have developed a link testing traceback technique that does not require any support from network operators [Burch 2000]. This technique can be called controlled flooding because it tests links by flooding them with large bursts of traffic and observing how this perturbs traffic from the attacker. Using a pregenerated "map" of Internet topology, the victim coerces selected hosts along the upstream route into iteratively flooding each incoming link on the router closest to the victim. Since router buffers are shared, packets traveling across the loaded link – including any sent by the attacker – have an increased probability of being dropped.

By observing changes in the rate of packets received from the attacker, the victim can therefore infer which link they arrived from.

### 3.3 Logging

An approach suggested in [Sager 1998] and [Stone 1999] is to log packets at key routers and then use data mining techniques to determine the path that the packets traversed. This scheme has the useful property that it can trace an attack long after the attack has completed (see Figure (1)). In Figure (1) network as seen from the victim of an attack, V. Routers are represented by Ri, and potential attackers by Ai. The dotted line represents a particular attack path between an attacker and the victim [Savage 2000]. However, it also has obvious drawbacks, including enormous resource requirements and a large-scale inter-provider database integration problem.

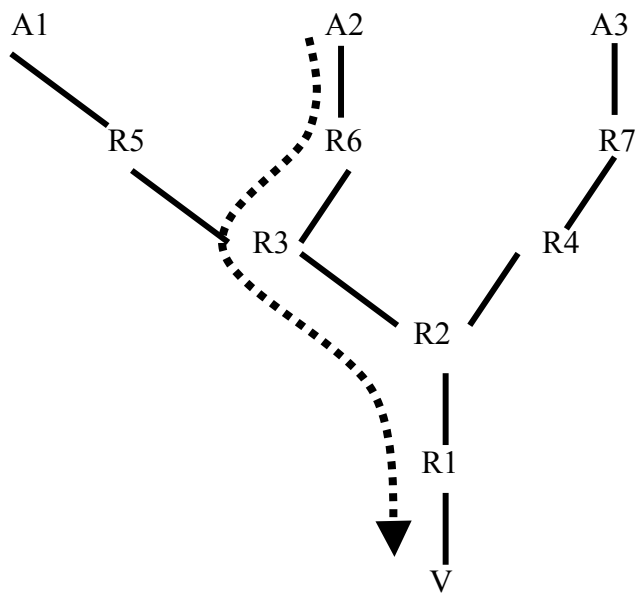

**Figure (1)**: Network attack path between an attacker and the victim [Savage 2000].

**Probabilistic Packet Marking (PPM)** is a traceback algorithm [Savage 2000]. It is motivated by the increased frequency and sophistication of denial-of-service attacks and by the difficulty in tracing packets with incorrect, or "spoofed", source addresses. This approach allows a victim to identify the network path(s) traversed by an attacker without requiring interactive operational support from Internet Service Providers (ISPs). Moreover, this traceback can be performed "post-mortem" – after an attack has completed. *In the proposed model we suggest use the PPM algorithm for denial-of-service attacks because it is compatible and can be efficiently implemented using conventional technology.*

## 4. IP Tracing and Active Response Model

The proposed model, as such, exploits technologies and techniques already in existence. It provides a suitable infrastructure that can perform rapid response to attacks. Moreover it uses different tracing approach to identify the real IP address for the attacking host. This model consists of both Passive Protection Mechanism (PPM) and Active Response Mechanism (ARM) integrated together in one model as shown in Figure (2).

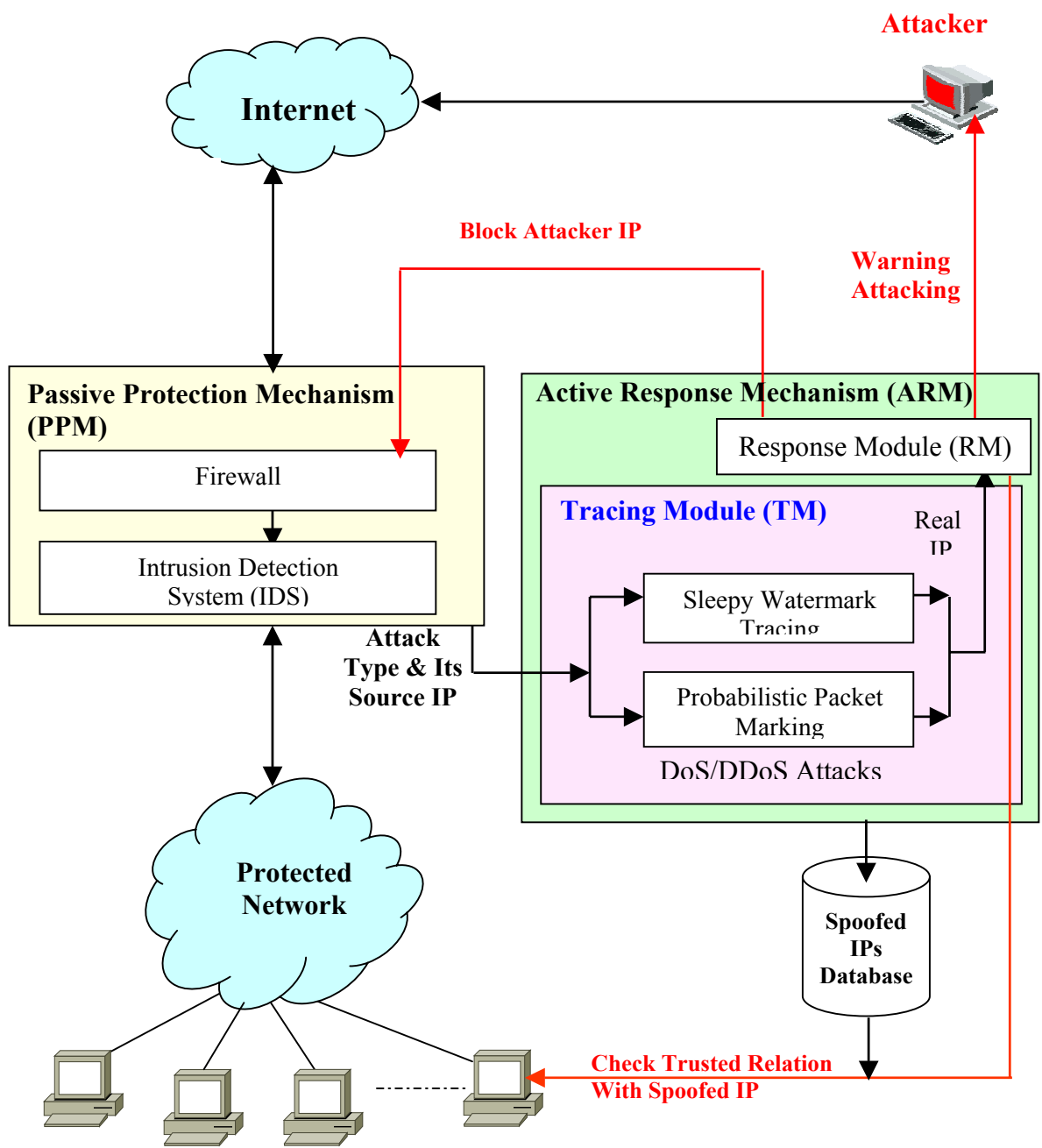

**Figure 2:** IP Tracing and Active Response Model

## 4.1 Passive Protection Mechanism

Detecting the attacks without taking any actions is called "Passive Network Protection". It takes many different forms such as firewall, access control and intrusion detection systems.

A Firewall provides controlled access between a private network and the Internet. It intercepts each message between the private network and the Internet. Depending on the configuration, the firewall determines whether the data packet or a connection request should be permitted to pass through the firewall or be discarded. But there are several types of security exposures to private networks that a firewall cannot address:

- Internal Intrusion: the firewall cannot protect the resources from attack by an internal user of the private networks.
- Direct Internet traffic: The firewall cannot protect he resources of the private network from the traffic that takes place directly with the Internet.
- Virus Protection: A firewall cannot protect a private network from an external virus. A virus may be transferred to the private network using (FTP) file transfer protocol or other means.

The best intrusion prevention system will fail. A system second line of defense is intrusion detection. Intrusion detection is identified using the following approaches:
- Statistical Anomaly Detection
- Rule-Based Detection

PPM based on two popular protection systems firewall and intrusion detection. Firewall used as first line of defense to filter both incoming and outgoing packets according to organization policy. Intrusion detection system used as a second line of defense in order to detect both external intruder from outside the underlying network and internal intruder inside underlying network.

## 4.2 Active Response Mechanism

Recently the computer user has become far better informed than ever, but as networks get bigger and bigger more naive users are introduced. Some of them are unwittingly causing many security gaps. At the same time, today's the hacker is more well informed too, and increasingly able to take advantage of those gaps and get inside the target network.

Therefore, we have to be diligent in cleaning up the security holes as they arise. Unfortunately, the increasing burden of maintaining network security is squeezing already overburdened system administrators. Existing *passive* security analysis tools generate too much data that takes too much work to parse.

The fundamental problem with passive security trends is the huge amount of required computations. Because such systems passively monitor the network traffic, they have to record all the concurrent incoming and outgoing connections even when there is no intrusion to trace. To correlate relevant events at any host they need to match every incoming connection with every outgoing connection of that host. Thus, for a host with $m$ concurrent incoming connections and $n$ concurrent outgoing connections, the passive intrusion detection tool would take $O(m \times n)$ comparisons, in addition to the $O(m+n)$ recording of concurrent connections. Moreover, it needs a further human decision and a manual action, such as running a firewall program to shut down a particular network port.

Active Response Mechanism (ARM) contains two modules Tracing Module (TM) and Response Module (RM).

TM in this work consists of two cooperative tracing modules. First is Sleepy Watermark Tracing (SWT). Second is Probabilistic Packet Marking (PPM) algorithm. In SWT module, intrusions are those attacks aiming to gain unauthorized access, rather than denial of service attacks. The compromised router needs to be addressed first, before the tracing of the intrusion can go any further. In this case SWT can still trace to the farthest trustworthy guardian gateway.

PPM algorithm used for denial-of-service attacks because it is compatible and can be efficiently implemented using conventional technology. This class of algorithm, best embodied in edge sampling, can enable efficient and robust multi-party traceback that can be incrementally deployed and efficiently implemented.

TM archive spoofed (masked) IPs into a database. Spoofed (masked) IPs database used to improve tracing algorithms when the same attack source IP comes back.

RM receives real attacker source address from TM. RM is responsible for taken appropriate action against attacker. Such actions may range from just warning to blocking his traffic or even attacking some of his resources. Some times these actions may be firewall reconfiguration such as deny access for certain IP or range of IPs or closing certain port like FTP port. Also, some times these actions may be changing file or system or account permissions.

## 5. Conclusion

The concept of active security that exploits active networking principles can offer a wide range of opportunities to build better security systems. In this paper, we have argued that existing tracing intrusion defense approaches. We have proposed a more active response paradigm to help to better repel or eliminate network-based intrusions and have identified the need of effective network-wide intrusion source tracing in order to build automated, network-wide response system.

The proposed model can do the following tasks:
- An active response mechanism to afford appropriate actions against the real intruder. Such actions may range from just warning to blocking his traffic or even attacking some of his resources.
- Protection of other network targets from being victimized by the same intruder.

This paper presented SWT [Wang 2001] as an example of network-wide intrusion tracing capability built upon SWT to dynamically push the intrusion defense perimeter close to the source of network-based intrusion. Also, we have argued that denial-of-service attacks motivate the development of improved traceback capabilities and we have suggested traceback algorithms based on packet marking in the network [Savage 2000].

- SWT approach, intrusions are those attacks aiming to gain unauthorized access, rather than denial of service attacks.
- PPM algorithm for both Denial of Service and Distributed Denial of Service attacks.